\documentclass[aps,prl,showpacs,twocolumn,groupedaddress]{revtex4}
\usepackage{amsmath,amsthm,amssymb}
\usepackage[dvips]{graphicx}

\begin{document}

\newcommand{\bi}[1]{\ensuremath{\boldsymbol{#1}}} 
\def\bra#1{\langle #1}
\def\ket#1{#1 \rangle}
\def\braket#1{\langle #1\rangle}
\def\Rnum#1{\uppercase\expandafter{\romannumeral #1}} 

\title{Perturbation on Hyperfine-enhanced $^{141}$Pr Nuclear Spin Dynamics 
Associated with Antiferroquadrupolar Order in PrV$_2$Al$_{20}$}

\author{T.~U.~Ito$^{1,2}$} 
\thanks{ito.takashi15@jaea.go.jp}
\author{W.~Higemoto$^{1,2,3}$} 
\author{A.~Sakai$^{4}$}
\author{M.~Tsujimoto$^{4}$}
\author{S.~Nakatsuji$^{4}$}
\affiliation{$^1$Advanced Science Research Center, Japan Atomic Energy
 Agency, Tokai, Ibaraki 319-1195, Japan}
\affiliation{$^2$J-PARC Center, Japan Atomic Energy Agency, Tokai,
Ibaraki 319-1195, Japan}
\affiliation{$^3$Department of Physics, Tokyo Institute of Technology, 
Meguro, Tokyo 152-8551, Japan}
\affiliation{$^4$Institute for Solid State Physics, University of Tokyo,
Kashiwa, Chiba 277-8581, Japan}


\begin{abstract}
The nature of multipolar order and hyperfine-enhanced (HE) 
$^{141}$Pr nuclear spin dynamics in PrV$_2$Al$_{20}$ was 
investigated using the muon spin relaxation technique. 
No explicit sign of time-reversal symmetry breaking was found below the 
multipolar order temperature $T_Q\sim 0.6$~K in a zero applied field as
anticipated on the basis of the antiferroquadrupolar (AFQ) order picture
proposed by Sakai and Nakatsuji [J.~Phys.~Soc.~Jpn. {\bf 80}, 063701 (2011)]. 
Further evidence of the nonmagnetic ground state was obtained from the
 observation of HE $^{141}$Pr nuclear spin fluctuations in the MHz scale. 
A marked increase in the muon spin-lattice relaxation rate
(1/$T_{\rm 1,\mu}$) was observed below 1~K with decreasing
temperature, which was attributed to the perturbation on the HE
$^{141}$Pr nuclear spin dynamics associated with the development
 of AFQ correlations. The longitudinal field dependence of 1/$T_{\rm
 1,\mu}$ revealed that the enhanced $^{141}$Pr nuclear spin accidentally
 has an effective gyromagnetic ratio close to that of the muon. 
\end{abstract}



\maketitle
\section{\Rnum{1}. INTRODUCTION}
Recently, considerable attention has been paid to the quadrupolar
degrees of freedom (DOF) of $4f$-electrons in Pr-based compounds with
the non-Kramers $\Gamma_3$ crystalline-electric-field (CEF) ground
doublet. Various novel phenomena
related to $\Gamma_{3g}$ quadrupoles,
such as incommensurate quadrupolar order, multi-channel Kondo
effects, quadrupolar quantum criticality, and consequent heavy fermion
superconductivity, have been intensively studied~\cite{Onimaru05,Yatskar96,Tanida06,Sakai11,Matsubayashi12,Tsujimoto14,Onimaru11}. However,
experimental techniques to probe quadrupolar properties are still
quite limited. The development of new methodologies is critical for
the further advancement of this research field.

The hyperfine enhancement of $^{141}$Pr nuclear magnetism is a common
phenomenon for $\Gamma_1$ and $\Gamma_3$ CEF ground multiplets without
dipolar DOF~\cite{Bleaney73,Abe03,Ito09,Iwakami14}. 
This effect arises from the Van Vleck-like admixture of magnetic CEF excited
multiplets into the nonmagnetic ground multiplets as a result of strong
intra-atomic hyperfine coupling~\cite{Bleaney73}. The $^{141}$Pr nuclear
spin-spin interaction is mediated by electronic exchange between
hyperfine-induced $4f$ moments. Therefore, the $4f$ quadrupolar state in
the $\Gamma_3$
ground doublet can potentially be probed via hyperfine-enhanced (HE)
$^{141}$Pr nuclear spin dynamics.

In this paper, we report an observation of
quadrupole-induced perturbation on HE $^{141}$Pr nuclear spin dynamics
in the $\Gamma_3$ ground doublet system PrV$_2$Al$_{20}$ using the
muon spin relaxation ($\mu$SR) technique.
PrV$_2$Al$_{20}$ shows multipolar order at $T_Q\sim 0.6$~K, which is well
below the temperature corresponding to the first excited CEF level at
$\Delta_{ex}/k_B\sim 40$~K~\cite{Sakai11}.
The primary order parameter is supposed to be a $\Gamma_{3g}$ quadrupole
based on active multipolar DOF in the $\Gamma_3$ ground doublet, entropy
release $<R$ln2, and magnetization~\cite{Sakai11,Shimura13}. These are
similar to those in isostructural PrTi$_2$Al$_{20}$
($T_Q\sim 2.0$~K~\cite{Sakai11},
$\Delta_{ex}/k_B\sim 65$~K~\cite{Sato12}); however, the field dependences
of the specific heat anomalies at $T_Q$ are totally different. The width
of the specific heat peak becomes broader with increasing field in
PrTi$_2$Al$_{20}$, whereas it is 
almost field-independent in PrV$_2$Al$_{20}$~\cite{Sakai11}. These
responses to applied magnetic fields suggest ferro- and
antiferro-quadrupolar (FQ and AFQ) order in Ti and V compounds, respectively. 
In PrTi$_2$Al$_{20}$, the FQ order has
been definitely identified from microscopic points of view using
$\mu$SR, NMR, and neutron scattering
techniques and the primary order parameter has been determined to be an
$O_2^0$-type $\Gamma_{3g}$ quadrupolar moment~\cite{Ito11,Tokunaga13,Sato12}.
By contrast, no direct microscopic evidence of the putative AFQ
order in PrV$_2$Al$_{20}$ has been provided to date. Herein, we first establish
the nonmagnetic nature of the primary order parameter in
PrV$_2$Al$_{20}$ from the $\mu$SR point of view using its high sensitivity
to local magnetic fields. This provides a strong justification for the AFQ order 
and AFQ quantum criticality at ambient
pressure~\cite{Sakai11,Tsujimoto14,Shimura15}.
Next, we show that the muon spin-lattice relaxation rate (1/$T_{\rm
1,\mu}$) exhibits a step-like change at around
$T_Q$, which can be attributed to the perturbation on the strength of
electron-mediated $^{141}$Pr nuclear spin interactions.
A comparison is made with the flat temperature dependence of 
$1/T_{1,\mu}$ reported for the FQ compound PrTi$_2$Al$_{20}$~\cite{Ito11}.

\section{\Rnum{2}. EXPERIMENTAL}
Single-crystalline samples of PrV$_2$Al$_{20}$ were prepared by the Al
self-flux method~\cite{Sakai11}. Pulsed $\mu$SR measurements were
performed under a zero applied field (ZF) and longitudinal magnetic fields
($B_0$) at the D1 area of
the J-PARC muon facility, Tokai, Japan, using the D$\Omega$1 spectrometer.
$\mu$SR spectra were recorded over the temperature ranges of 0.045-3~K and
3-40~K with a $^3$He-$^4$He dilution refrigerator and a conventional
$^4$He flow cryostat, respectively. The PrV$_2$Al$_{20}$ single crystals were
randomly aligned and glued on silver sample holders with commercial
Apiezon N grease. Spin-polarized single-bunch muon pulses were incident
on the samples with initial muon spin polarization \bi{P}$(t=0)$
antiparallel to the beam incident direction. $\mu$-decay positrons
were detected by forward and backward positron counters.
Because our samples do not show any sign of superconductivity down to 0.045~K,
a possibility of time-reversal symmetry breaking associated with
superconductivity can be ignored.

\section{\Rnum{3}. RESULTS AND DISCUSSION}
\begin{figure}[tb]
\begin{center}
\includegraphics[scale =0.54]{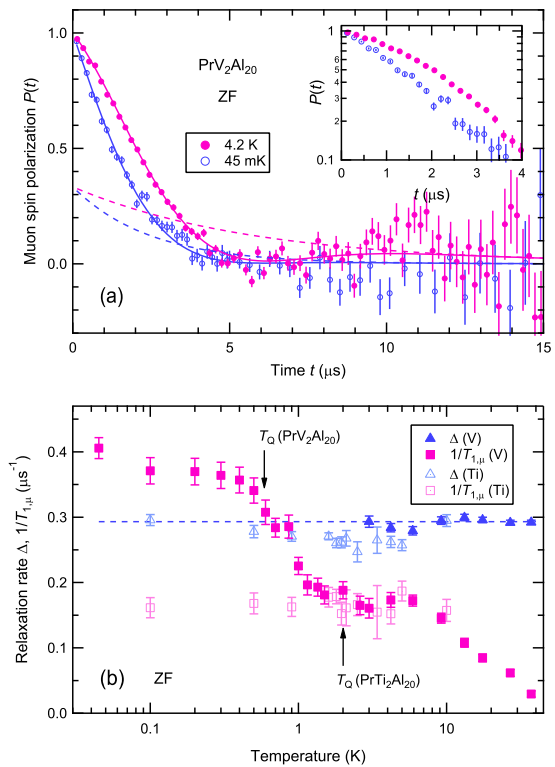}
\end{center}
\caption{(a) ZF-$\mu$SR spectra of PrV$_2$Al$_{20}$ at 4.2~K
 (closed circles) and 45~mK (open circles). The solid curves represent
 the best fits to
 Eq.(\ref{eq1}). The broken curves designate the 1/3 component. The
 inset shows the same data with $P(t)$ plotted on a
 log scale. (b) $\Delta$ (triangles) and $1/T_{1,\mu}$ (squares) under
 ZF in PrV$_2$Al$_{20}$ (closed symbols) and PrTi$_2$Al$_{20}$ (open
 symbols, data from Ref.\cite{Ito11}) as
 functions of temperature. The broken line represents the average value
 of $\Delta$ for PrV$_2$Al$_{20}$.} 
\label{f1}
\end{figure}

Figure~\ref{f1}(a) shows the ZF-$\mu$SR spectra of PrV$_2$Al$_{20}$ at
4.2 and 0.045~K. $P(t)$ is the projection of \bi{P}$(t)$ onto the
beam incident axis and has been normalized after subtracting the background
signal from the silver sample holders. $P(t)$ at 4.2~K above
$T_Q$ exhibits a Gaussian-like damping in the early-time region
and a slight recovery after 6~$\mu$s. These features can be modeled 
well with the function
\begin{align}
P(t) &= e^{-t/T_{1,\mu}}G_{KT}(t;\Delta,B_0=0) \nonumber \\
&=\frac{1}{3}e^{-t/T_{1,\mu}}+\frac{2}{3}(1-\Delta^2
 t^2) e^{-\frac{1}{2}\Delta^2 t^2 -t/T_{1,\mu}}, \label{eq1}
\end{align}
where the exponential function describes $T_1$ relaxation caused by
magnetic fluctuations, and the static Gaussian Kubo-Toyabe function
$G_{KT}(t;\Delta,B_0)$ with the relaxation rate $\Delta$ expresses loss of
muon spin coherence under static local fields with an isotropic Gaussian
probability distribution~\cite{Hayano79}.
This model was also adopted in Ref.\cite{Ito11} to describe 
ZF-$\mu$SR in PrTi$_2$Al$_{20}$, where the primary origins of the
fluctuating and static local fields were determined to be HE $^{141}$Pr
and bare $^{27}$Al nuclear spins, respectively.
These imply that a similar mechanism is also in effect in
PrV$_2$Al$_{20}$.

The damping of the spectrum at 0.045~K is obviously faster than that at
4.2~K. Supposing that the additional damping were entirely due to the
development of static local fields below $T_Q$, the extra field spread
would be roughly estimated to be $(\tau_{0.045{\rm
K}}^{-2}-\tau_{4.2{\rm K}}^{-2})^{1/2}/\gamma_{\mu}
\sim 4\times10^{-4}$~T, where $\tau$ is the 1/$e$ width, and
$\gamma_{\mu}$ ($=2\pi\times 135.53$~MHz/T) is the muon gyromagnetic ratio. 
On the other hand, muons in SmTi$_2$Al$_{20}$ with a 
0.51-$\mu_B$/Sm ordered moment feel a local field of
$\sim 5 \times10^{-2}$~T~\cite{Higashinaka11,Ito12}. From a simple scaling,
the magnitude of the hypothetical ordered moment in
PrV$_2$Al$_{20}$ is estimated to be $4\times10^{-3}~\mu_B$/Pr. This is
too small to be associated with the
entropy release $\sim 0.5R$ln2 at $T_Q$~\cite{Sakai11}. 
Therefore, the possibility of magnetic order and consequent
development of static local fields is ruled out in
PrV$_2$Al$_{20}$. The ZF spectrum at 0.045~K is more exponential-like in
shape as shown in the inset of Fig.~\ref{f1}(a). This suggests that the
additional damping is primarily due to an increase in
1/$T_{\rm 1,\mu}$. Further evidence can be obtained by carefully
investigating the 1/3 component as the first term in Eq.(\ref{eq1}). The
relaxation of this component is caused by the $T_1$ process under
effective longitudinal fields associated with the longitudinal component
of the static nuclear dipolar fields along
\bi{P}$(t=0)$~\cite{Hayano79}. Therefore, the loss of the recovery after
6~$\mu$s at 0.045~K manifests the increase in 1/$T_{\rm 1,\mu}$. 
From our ZF-$\mu$SR measurements, no explicit proof of time-reversal
symmetry breaking was found below $T_Q$. This strongly
suggests that the order parameter is a time-reversal-even multipole, supporting
the AFQ order scenario from a microscopic point of view. Note that a
$T_{xyz}$-type magnetic octupole is also active in the $\Gamma_3$
subspace~\cite{Tsujimoto14}. Our results suggest that $T_{xyz}$
octupolar order is unlikely in our samples.

ZF-$\mu$SR spectra were fit to Eq.(\ref{eq1}) to extract the
temperature dependences of $\Delta$ and $1/T_{1,\mu}$.
First, fits were performed in the entire temperature range with
$\Delta$ and $1/T_{1,\mu}$ being free. The values of $\Delta$ obtained
from the fits were almost constant above 3~K. This is reasonable because
$\Delta$ resulting from the $^{27}$Al and $^{51}$V nuclear dipolar
moments is expected to be independent of temperature in the temperature
range where muons are immobile.  
The uncertainty in $\Delta$ steeply
increases below 3~K as $T_1$ relaxation becomes dominant. This hinders
the precise estimation of $1/T_{1,\mu}$ at low temperatures; therefore,
we fixed $\Delta$ to the average value above 3~K and fit the spectra
below 3~K with only $1/T_{1,\mu}$ being free. Satisfactory fits were
obtained, as shown by the solid curves in Fig.~\ref{f1}(a).

The values of $\Delta$ and $1/T_{1,\mu}$
for PrV$_2$Al$_{20}$ are shown by the solid triangles and
squares, respectively, in Fig.~\ref{f1}(b). 
Those for PrTi$_2$Al$_{20}$ from Ref.\cite{Ito11}
are also plotted with corresponding open symbols.
The $\Delta$ values of both compounds are in good agreement, further
demonstrating the validity of our model and fitting procedure for
PrV$_2$Al$_{20}$.
The root-mean-square (rms) width of the Gaussian local field distribution
$\Delta/\gamma_{\mu}\sim 3.5 \times10^{-4}$~T is reasonable for
abundant $^{27}$Al and $^{51}V$ nuclei~\cite{Ito12}. 
The $1/T_{1,\mu}$ of PrV$_2$Al$_{20}$ exhibits a double-plateau
structure, as shown in Fig.~\ref{f1}(b). An increase in $1/T_{1,\mu}$
with decreasing temperature in the temperature range of 6-40~K is
ascribed to the development of HE $^{141}$Pr nuclear moments associated with
the increased Van Vleck contribution in magnetic susceptibility. The first
plateau in the temperature range of 1-6~K suggests that
exchange-mediated $^{141}$Pr spin-spin interactions are fully developed
and the $^{141}$Pr nuclear spin fluctuation rate $\nu$ is consequently
temperature-independent.
In PrTi$_2$Al$_{20}$, this plateau extends down to 0.1~K without any
significant anomaly at $T_Q\sim2.0$~K.
By contrast, the $1/T_{1,\mu}$ of PrV$_2$Al$_{20}$ clearly increases
with decreasing temperature below 1~K, and a second plateau 
forms below $T_Q\sim0.6$~K. This behavior suggests that the
exchange-mediated $^{141}$Pr nuclear spin interactions are effectively
weakened as AFQ correlations develop below 1~K. 
The significant difference between the FQ and AFQ
compounds implies that the antiferro-type correlation might be essential
for this perturbation. 

Note that low energy magnetic excitations in a
magnetically ordered state can also contribute to the $T_1$
relaxation. When this process is dominant, however, $1/T_{1,\mu}$ should steeply
decrease with decreasing temperature as the low energy excitations are
suppressed. This is clearly not the case in PrV$_2$Al$_{20}$, and
therefore this possibility is excluded.

One may associate the difference between $T_Q$ and the onset temperature of
the increase in $1/T_{1,\mu}$ with a possibility of muon-charge-induced
nucleation of a quadrupolar cluster slightly above $T_Q$. Unfortunately, it
is difficult to completely rule out such a possibility from our
data. However, even if that is the case, the sharp contrast between
the FQ and AFQ compounds still suggests the importance of AFQ
correlations for understanding the behavior of $1/T_{1,\mu}$ in
PrV$_2$Al$_{20}$.

\begin{figure}[tb]
\begin{center}
\includegraphics[scale =0.65]{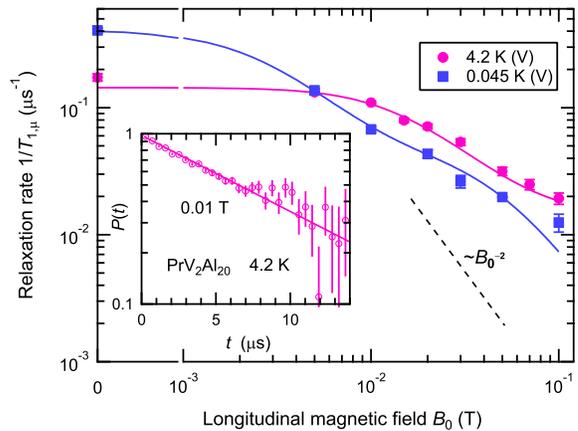}
\end{center}
\caption{$B_0$-dependences of $1/T_{1,\mu}$ at 4.2~K
 (circles) and 0.045~K (squares) in PrV$_2$Al$_{20}$. The horizontal
 axis is on a linear scale for $B_0< 10^{-3}$~T and on a
 log scale for $B_0\geq 10^{-3}$~T. The solid curves represent the best fits
 to Eq.(\ref{eq2}).
The broken line illustrates the slope of functions that follow a
 $B_0^{-2}$ dependence.
The inset shows the $\mu$SR spectrum at 4.2~K in $B_0=0.01$~T.
}
 \label{f2}
\end{figure}
Figure~\ref{f2} shows the $B_0$-dependences of $1/T_{1,\mu}$ in
PrV$_2$Al$_{20}$ at 4.2 and 0.045~K.
The $1/T_{1,\mu}$ value for $B_0>0$ was obtained from fits to
$P(t)=e^{-t/T_{1,\mu}}G_{KT}(t;\Delta,B_0)$ with $\Delta$ being fixed to
the average value in ZF. The validity of our single-$T_1$ model can be
visually checked in $B_0\geq 0.01$~T, where $G_{KT}(t;\Delta,B_0)\sim
1$ holds. All spectra above 0.01~T follow a single-exponential function
well, as the example shown in the inset of
Fig.~\ref{f2} illustrates. The influence of avoided level crossing
resonance~\cite{Kreitzman86} with $^{27}$Al seems negligible since the
$1/T_{1,\mu}$-$B_0$ curves show smooth changes without any significant anomaly.

The $1/T_{1,\mu}$ owing to the dipolar
coupling between HE $^{141}$Pr nuclear and muon spins in $B_0$ is described by 
\begin{align}
\frac{1}{T_{1,\mu}} &= \frac{\sigma_{B}^2\gamma_{\mu}^2}{5}
\biggl\{\frac{3\nu}{\nu^2+\gamma_{\mu}^2B_0^2}\nonumber \\
&+ \frac{\nu}{\nu^2+[\gamma_{\mu}-\gamma_{I}^*]^2B_0^2} 
+ \frac{6\nu}{\nu^2+[\gamma_{\mu}+\gamma_{I}^*]^2B_0^2} \biggr\},\label{eq2}
\end{align}
where $\sigma_{B}$ and $\gamma_{I}^*$ are the rms width of the local
field distribution and the effective gyromagnetic ratio for the HE
$^{141}$Pr nuclear spin, respectively~\cite{Hayano79,Shu07}. $\gamma_{I}^*$ is
enhanced by a factor of $(1+K)$ compared to the bare $^{141}$Pr gyromagnetic
ratio ($\gamma_{I}=2\pi\times 13.054(2)$~MHz/T~\cite{Stone05}), where
$K$ is the $^{141}$Pr Knight shift.
As an approximation, we use an orientation-averaged $K$ in the AFQ ordered
state, where anisotropy in $K$ is expected to arise because of the
splitting of the $\Gamma_3$ doublet.
A simpler form of Eq.~(\ref{eq2}) with $\gamma_I^*=0$ is frequently
used, as was adopted in Ref.~\cite{Ito11} for fitting
$1/T_{1,\mu}(B_0)$ of PrTi$_2$Al$_{20}$.
When $\gamma_{I}^*$ is comparable with $\gamma_{\mu}$ (namely, $K \sim 9.4$), the second term in Eq.(\ref{eq2}) results in a high-field tail in
the plot of $1/T_{1,\mu}$ versus $B_0$. 
This should be the case in PrV$_2$Al$_{20}$ because $K$ is roughly estimated to
be 12 by the relationship $K=a_{hf}\chi_{4f}$, where
$a_{hf}$ ($=187.7$~mol/emu~\cite{Andres77}) is the hyperfine coupling
constant for Pr$^{3+}$ and $\chi_{4f}$ ($=0.067$~emu/mol at
2~K~\cite{Sakai11}) is the molar $4f$ susceptibility.
Fits to Eq.(\ref{eq2}) were performed without restraints on
$\sigma_{B}$, $\nu$, and $K$.
Satisfactory fits were obtained, as shown by the solid curves in
Fig.~\ref{f2}. 

\begin{table}
\begin{ruledtabular}
\caption{$\nu$, $\sigma_B$, and $K$ for PrV$_2$Al$_{20}$ at 4.2 and 0.045~K.}
\label{t1}
\begin{tabular}{cccc}
$T$~(K)~&  ~$\nu$~(MHz)~ & ~$\sigma_B$~(10$^{-4}$~T)~ & ~$K$~\\
\hline
 4.2 & 22.7 $^{+1.4}_{-1.3}$ & 14.9$\pm$0.4 & 9.8$^{+0.6}_{-0.7}$\\
 0.045 & 4.0$\pm$0.2 & 10.5$\pm$0.2 & 10.4$\pm$0.1 \\
\end{tabular}
\end{ruledtabular}
\end{table}
The fitting parameters for 4.2 and 0.045~K are listed in
Table~\ref{t1}. 
The MHz-scale $\nu$ is typical of exchange-mediated $^{141}$Pr
nuclear spin-spin interactions in nonmagnetic CEF ground
states~\cite{Macl00,Aoki03,Shu07,Tokunaga10,Ito11,Ito09},
further justifying our model. A marked reduction in $\nu$ 
 at 0.045~K clarifies that the step-like increase in $1/T_{1,\mu}$ below
 1~K in ZF is mainly due to the slowing down of $^{141}$Pr nuclear spin
 fluctuations. Taking $\nu$ at 0.045~K as a measure of the effective nuclear
 exchange constant $|J_{nucl}|/\hbar$ in the ground state, we can
 estimate the $^{141}$Pr nuclear order temperature $T_{NO}$ using 
 the following relationship: $T_{NO}=|J_{nucl}|I(I+1)/3k_B$, where
 $I=5/2$ is the $^{141}$Pr nuclear spin. Accordingly, $T_{NO}$ for
 PrV$_2$Al$_{20}$
 is estimated to be 89(5)~$\mu$K, slightly lower than that estimated for
 PrTi$_2$Al$_{20}$~\cite{Ito11}.
The value of $\sigma_{B}$ is significantly larger than
$\Delta/\gamma_{\mu}\sim 3.5 \times 10^{-4}$~T associated with 
$^{27}$Al and $^{51}$V nuclei. Together with the large $K$,
this is consistent with the hyperfine enhancement picture. Such an
effect occurs only when the Pr$^{3+}$ ground state does not involve
active dipolar DOF~\cite{Macl00}. Therefore, our
observation of the HE $^{141}$Pr nuclear spin dynamics provides further
microscopic evidence of the nonmagnetic $\Gamma_3$ ground doublet and 
AFQ order in PrV$_2$Al$_{20}$. 
A slight decrease in $\sigma_{B}$ at the lower
temperature can likely be ascribed to a
change in the shape of the local field distribution because of the
anisotropy in $K$ expected in the AFQ ordered state.
The $K$ values at 0.045 and 4.2~K agree within the error.
This is reasonable because the splitting of the $\Gamma_3$
doublet does not change the orientation-averaged value of single-ion Van
Vleck susceptibility when the $\Gamma_3$ splitting is negligibly small
compared to $\Delta_{ex}$. 
The fit to the data at 0.045~K slightly deviates at 0.1~T, as shown in 
Fig.~\ref{f2}. This might be due to the anisotropy in $K$ below $T_Q$,
which is not explicitly taken into account in Eq.(\ref{eq2}).

The temperature dependence of $1/T_{1,\mu}$ for
PrV$_2$Al$_{20}$ shown in Fig.~\ref{f1}(b) indicates that
$\nu$~($\propto|J_{nucl}|)$ begins to
decrease below 1~K and levels off at around $T_Q$. 
This behavior suggests that the strength of the $^{141}$Pr nuclear spin
coupling is effectively weakened as AFQ correlations develop. 
One possible origin for this reduction is intra-atomic electric
quadrupolar coupling between $4f$ and $^{141}$Pr quadrupolar moments.
Here we assume that the $4f$ ground state is one of the eigenstates for
the $O^0_2$ quadrupolar operator. Six-fold degenerate
$^{141}$Pr spin wavefunctions split into $|I_z$=$\pm
1/2\rangle$, $|\pm 3/2\rangle$, and $|\pm 5/2\rangle$ under the
electric field gradient arising from the on-site $O_2^0$ moment. 
These levels are separated by $h\nu_Q$ ($\pm 1/2\leftrightarrow \pm
3/2$) and $2h\nu_Q$ ($\pm 3/2\leftrightarrow \pm 5/2$). Following the
treatment in Ref.\cite{Ikushima98}, we estimated $\nu_{Q}$ to be
1.6~MHz using the radial $\langle r^{-3}\rangle _{4f}=5.369~a_0^{-3}$,
the Sternheimer factor $R_{4f}=0.1308$~\cite{Sternheimer66}, and the
$^{141}$Pr quadrupolar moment $Q=-0.059$ barn~\cite{Stone05}. 
The estimated $\nu_Q$ is not negligible compared with the unperturbed
$\nu$ at 4.2~K; thus, the intra-atomic quadrupolar coupling can
significantly reduce the transition probabilities between the separated levels. 

Other possible origins for the effective reduction in $|J_{nucl}|$
come from the path of the $^{141}$Pr nuclear spin exchange.
Considering that this is mediated by the
Ruderman-Kittel-Kasuya-Yoshida interactions between $4f$ dipolar
moments induced by the intra-atomic hyperfine interaction,
the orientation-averaged $|J_{nucl}|$ can be approximately expressed as
\begin{equation}
|J_{nucl}|=\left( \frac{\gamma_I \hbar}{g_J\mu_B} \right)^2 |J_{ff}|\cdot
 {\rm Tr}[K_+K_-]/3,\label{eq3}
\end{equation}
where $g_J(=4/5)$ is the Land\'e $g$-factor for Pr$^{3+}$, $|J_{ff}|$ is the $4f$ exchange constant,
and $K_{\pm}$ are $^{141}$Pr Knight shift tensors for the two closest Pr ions.
This relation suggests that perturbation of Tr$[K_+K_-]$/3 and/or
$|J_{ff}|$ can be responsible for the reduction in $|J_{nucl}|$. Here we 
focus on the contribution from the Knight shift factor because any change in
$|J_{ff}|$ is expected to be relatively small.
Calculating the single-ion Van Vleck susceptibility for the
$O_2^0$-eigenstates yields the diagonal $K_{\pm}$ with
a set of principal values expressed as $(K\mp K_a, K\mp K_a, K\pm 2K_a$),
where $K_a$ is an anisotropic part. Consequently, Tr$[K_+K_-]/3$ is
evaluated to be $K^2-2K_a^2$, which is smaller than $K^2$ for the
paraquadrupolar state and thus is consistent with the reduced
$|J_{nucl}|$. A similar conclusion is also reached for $O_2^2$-eigenstates.

The intra-atomic quadrupolar coupling should also be in effect in
the FQ compound PrTi$_2$Al$_{20}$, which can decrease $\nu$. 
The flat temperature dependence of $1/T_{1,\mu}$ in PrTi$_2$Al$_{20}$
suggests that other contributions compensate for this ``$decoupling$'' effect. 
In the case of the $O_2^0$-type FQ order, the
Knight shift factor in Eq.(\ref{eq3}) is replaced with
Tr$[K^2_{\pm}]/3=K^2+2K_a^2$. The enhancement in $|J_{nucl}|$ because of
this factor may be a source of the compensation.

\section{\Rnum{4}. CONCLUSION}
$\mu$SR is sensitive to slow spin fluctuations in the MHz scale and thus
is appropriate for probing HE $^{141}$Pr nuclear spin
dynamics. In this study, we used $\mu$SR to demonstrate for the first
time that the AFQ correlations of $4f$ electrons can significantly
perturb the strength of the HE $^{141}$Pr nuclear spin-spin interaction
in PrV$_2$Al$_{20}$.
This paves the way for an alternative approach to investigate quadrupolar
correlations in Pr-based compounds using local spin probes via the
observation of HE $^{141}$Pr nuclear spin dynamics.

\section*{ACKNOWLEDGMENTS}
\begin{acknowledgments}
We thank the staff of J-PARC for facility operation and Y.~Tokunaga,
S.~Kambe, H.~S.~Suzuki, and Y.~Matsumoto for helpful discussions. 
This work was partly supported by 
Grants-in-Aid for Scientific Research (Grants No. 24710101 
and No. 25707030) and Program for Advancing Strategic 
International Networks to Accelerate the Circulation of 
Talented Researchers (Grant No. R2604) from the Japan 
Society for the Promotion of Science, and by Grants-in-Aid 
for Scientific Research on Innovative Areas (Grants No. 
23108002, No. 26108717, No. 15H05882, and No. 15H05883) 
from the Ministry of Education, Culture, Sports, Science, and 
Technology of Japan.

\end{acknowledgments}

\end{document}